\documentclass[onecolumn]{aastex631}
\usepackage{amsmath}

\usepackage{hyperref}

\newcommand{\be}{\begin{equation}}
\newcommand{\ee}{\end{equation}}

\shorttitle{Subsolar neutron star mergers from fragmenting collapsar disks}
\shortauthors{B.~D.~Metzger, L.~Hui, M.~Cantiello}

\begin{document}

\title{Fragmentation in Gravitationally-unstable Collapsar Disks and Sub-Solar Neutron Star Mergers}

\author[0000-0002-4670-7509]{Brian D.~Metzger}
\affil{Department of Physics and Columbia Astrophysics Laboratory, Columbia University, New York, NY 10027, USA}
\affil{Center for Computational Astrophysics, Flatiron Institute, 162 5th Ave, New York, NY 10010, USA} 

\author[0000-0002-4670-7509]{Lam Hui}
\affil{Department of Physics and Columbia Astrophysics Laboratory, Columbia University, New York, NY 10027, USA}

\author[0000-0002-8171-8596]{Matteo Cantiello}
\affil{Center for Computational Astrophysics, Flatiron Institute, 162 5th Ave, New York, NY 10010, USA} 
\affil{Department of Astrophysical Sciences, Princeton University, Princeton, NJ 08544, USA}

\begin{abstract}
Although stable neutron stars (NS) can in principle exist down to masses $M_{\rm ns} \approx 0.1M_{\odot}$, standard models of stellar core-collapse predict a robust lower limit $M_{\rm ns} \gtrsim 1.2M_{\odot}$, roughly commensurate with the Chandrasekhar mass $M_{\rm Ch}$ of the progenitor's iron core (electron fraction $Y_{e} \approx 0.5$). However, this limit may be circumvented in sufficiently dense neutron-rich environments ($Y_{e} < 0.5$) for which $M_{\rm Ch} \propto Y_{e}^{2}$ is reduced to $\lesssim 1M_{\odot}$. Such physical conditions could arise in the black hole accretion disks formed from the collapse of rapidly-rotating stars (``collapsars''), as a result of gravitational instabilities and cooling-induced fragmentation, similar to models for planet formation in protostellar disks. We confirm that the conditions to form sub-solar mass NS (ssNS) may be marginally satisfied in the outer regions of massive neutrino-cooled collapsar disks. If the disk fragments into multiple ssNS, their subsequent coalescence offers a channel for precipitating sub-solar mass LIGO/Virgo gravitational-wave mergers that does not implicate primordial black holes. The model makes several additional predictions: (1) $\sim $Hz frequency Doppler modulation of the ssNS-merger gravitational wave signals due to the binary's orbital motion in the disk; (2) at least one additional gravitational wave event (coincident within $\lesssim$ hours), from the coalescence of the ssNS-merger remnant(s) with the central black hole; (3) an associated gamma-ray burst and supernova counterpart, the latter boosted in energy and enriched with $r$-process elements from the NS merger(s) embedded within the exploding stellar envelope (``kilonovae inside a supernova'').
\end{abstract}


\section{Introduction}

The vast majority of neutron stars (NS) in nature are formed when the (usually predominantly, iron and nickel) core of a massive star undergoes gravitational collapse at the end of its nuclear burning evolution (e.g., \citealt{Burrows&Vartanyan21}). A minority of NS are also likely formed from the accretion-induced collapse of white dwarfs in compact binary systems (e.g., \citealt{Nomoto&Kondo91}). In both cases, the characteristic mass of the collapsing body is set by the \citet{Chandrasekhar31} value:
\be
M_{\rm Ch} \simeq 1.45 M_{\odot}\left(\frac{Y_{e}}{0.5}\right)^{2},
\label{eq:Mch}
\ee
where $Y_{e}$ is the electron fraction, normalized to the value $Y_{e} \approx 0.44-0.5$ for symmetric or nearly-symmetric nuclei such as $^{16}$O or $^{56}$Fe. In detail, the stellar core's mass at collapse differs from Eq.~\eqref{eq:Mch} due to various corrections (e.g., finite entropy, photodisintegration and electron-captures), as does the gravitational mass of the final NS (subject, e.g., to neutrino losses and the precise explosion mass-cut). Nevertheless, modern supernova simulations predict NS masses in a range $\approx 1.2-1.6M_{\odot}$ (e.g., \citealt{Sukhbold+16,Burrows+19,Ertl+20,Woosley+20}) comparable to $M_{\rm Ch}$. The lower bound of this range is roughly compatible with the $\simeq 1.17M_{\odot}$ secondary companion in the pulsar binary PSR J0453+1559 (\citealt{Martinez+15}), claimed to be the least massive NS known (e.g., \citealt{Suwa+18}, however see \citealt{Tauris&Janka19}). This stands in contrast to theoretically-allowed (i.e., dynamically stable) NS solutions, which extend down to masses $\approx 0.1M_{\odot}$ (e.g., \citealt{Lattimer&Prakash04}) an order of magnitude below what appears possible to create via ordinary stellar evolution.

Although no known instances of a NS weighing less than a solar mass have yet been confirmed (however, see \citealt{Doroshenko+22}), their potential existence in nature is of great interest. If sub-solar NS (ssNS) are created or otherwise end up in tight stellar binaries orbiting another compact object, their resulting gravitational wave-driven coalescence may manifest in the growing sample of compact object mergers (\citealt{Mandel&Broekgaarden22}). Based on their first three science runs, the Advanced LIGO/Virgo observatories have placed upper limits on the rate of merging sub-solar compact objects with masses $\approx 0.2-1M_{\odot}$ (\citealt{Abbott_subsolar+18,Abbott_subsolar_22,LVK_23}; however, see \citealt{Morras+23}). The discovery of sub-solar mass mergers would seemingly have major implications for fundamental physics, insofar as it would offer arguably the cleanest evidence for the existence of primordial black holes formed at the beginning of the universe (see \citealt{Carr+21} for a review).\footnote{In principle, tidal effects on the gravitational waveform can distinguish mergers of sub-solar mass black holes from very low-mass ssNS (\citealt{Silva+16,Bandopadhyay+23,Crescimbeni+24}).} Given these high stakes, one is motivated to consider any astrophysical scenarios for creating and merging ssNS, even speculative ones (e.g., \citealt{Popov+07}).

The majority of massive stars are believed to be rotating slowly at death, leading to NS birth following a neutrino-driven supernova explosion or to the formation of slowly-spinning black holes (e.g., \citealt{Fuller+19}) and comparatively dim electromagnetic counterparts (e.g., \citealt{Antoni&Quataert23}).  However, a small fraction $\lesssim 1\%$ of massive stars appear to be rotating much faster upon collapse \citep{Cantiello+07}, giving rise to the rare population of gamma-ray bursts (GRBs; \citealt{Woosley&Bloom06}). 
The ``collapsar'' model postulates that GRBs of the so-called long-duration variety are powered by the formation of a massive torus which orbits and feeds infalling stellar material onto a newly-formed black hole at rates of up to a solar mass per second or greater \citep{Woosley93,MacFadyen&Woosley99}.

The outer regions of massive collapsar accretion disks are susceptible to instabilities driven by self-gravity \citep{Chen&Beloborodov07}. In analogy with models for planet formation in protostar accretion disks (e.g., \citealt{Boss97,Lodato&Rice04,Chen+23}) or star formation in accretion disks around supermassive black holes (e.g., \citealt{Levin03,Goodman&Tan04}), cooling-induced fragmention in gravitationally-unstable collapsar disks offers the potential to form self-bound objects with masses comparable to NS. Such a scenario was proposed by \citet{Piro&Pfahl07}, motivated by the potential of such disk-formed NS to generate a detectable gravitational-wave signal upon coalescing with the central black hole. 

\begin{figure}
    \centering
\includegraphics[width=0.55\textwidth]{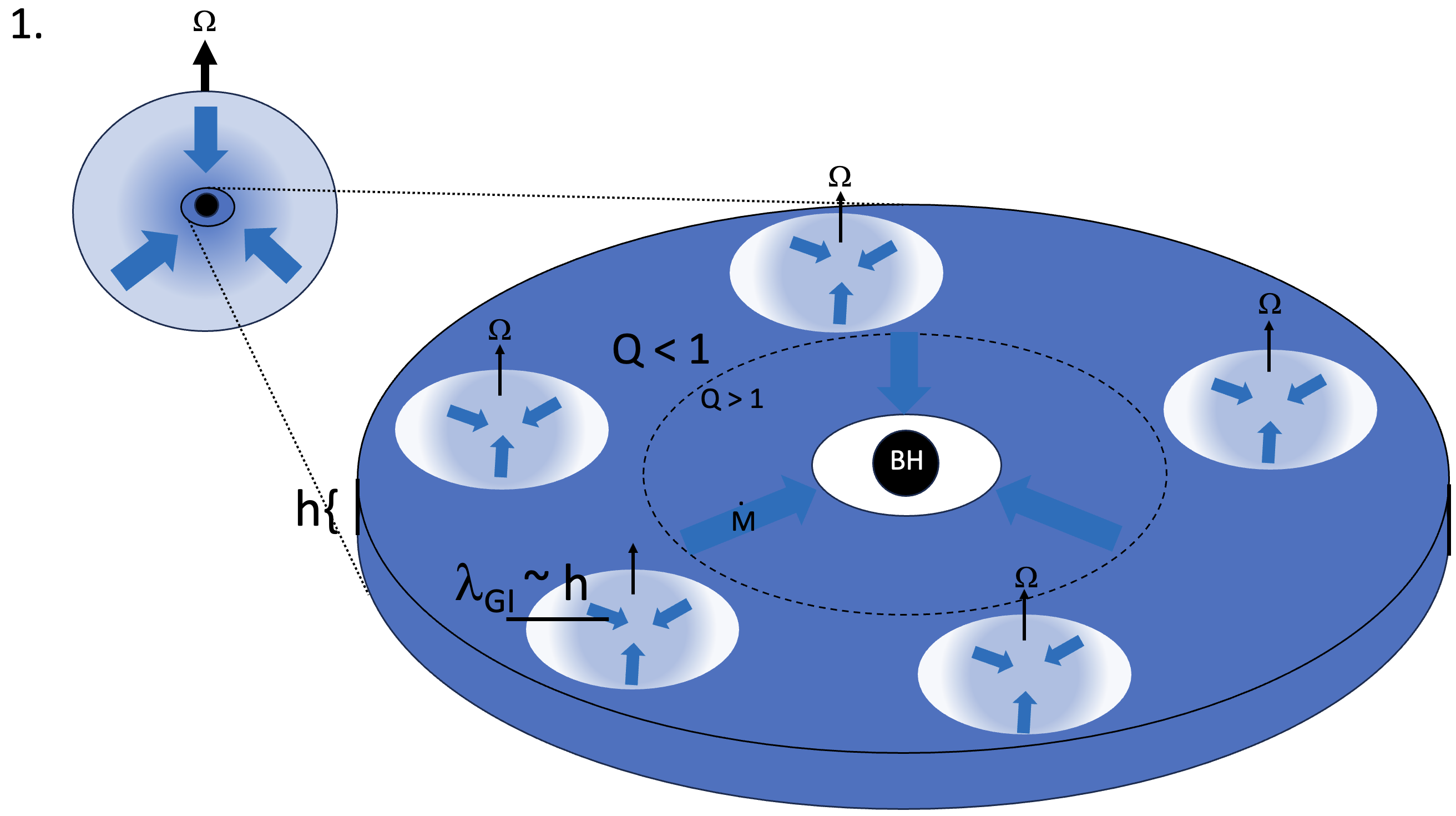}
\includegraphics[width=0.6\textwidth]{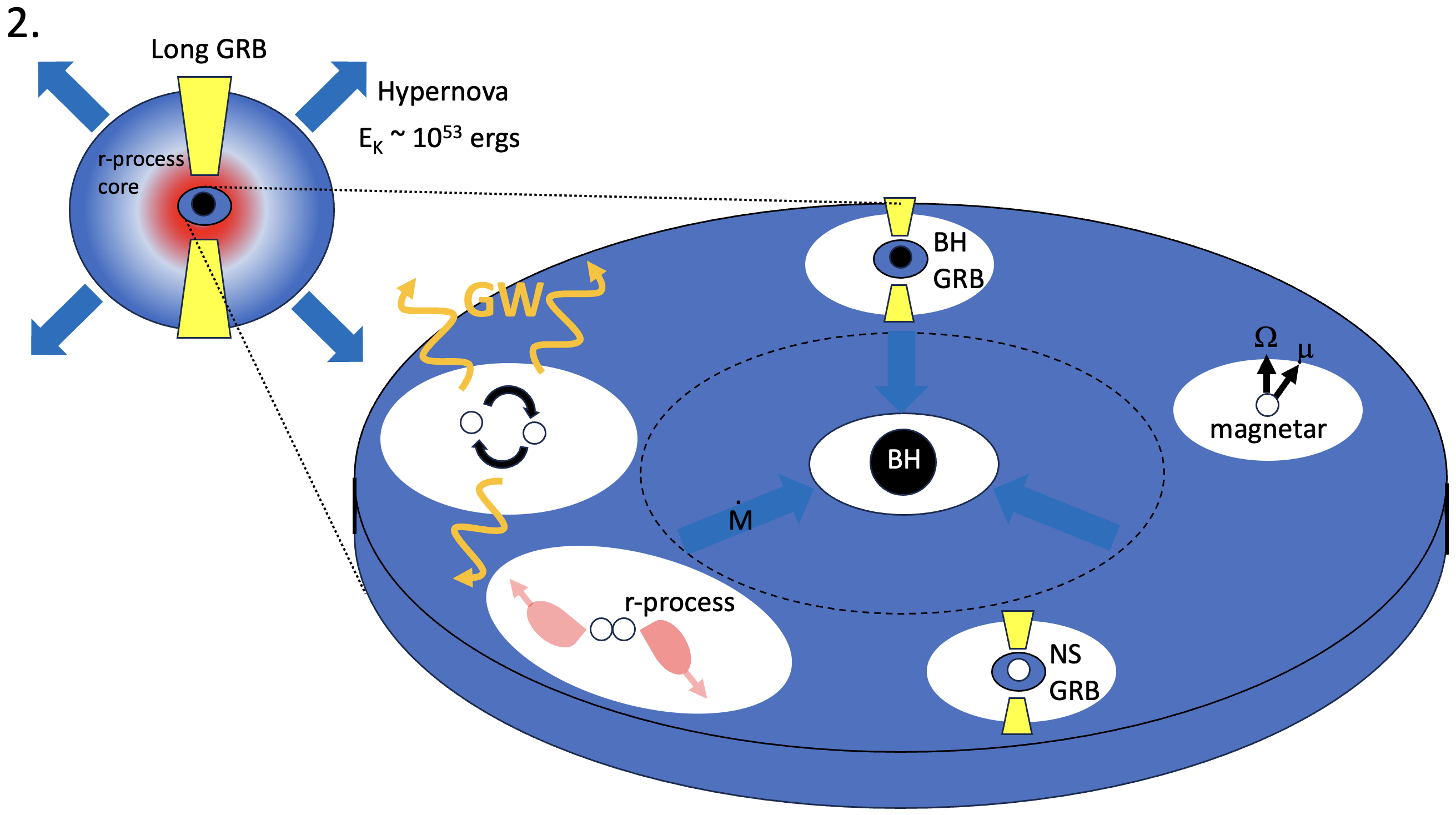}
\includegraphics[width=0.6\textwidth]{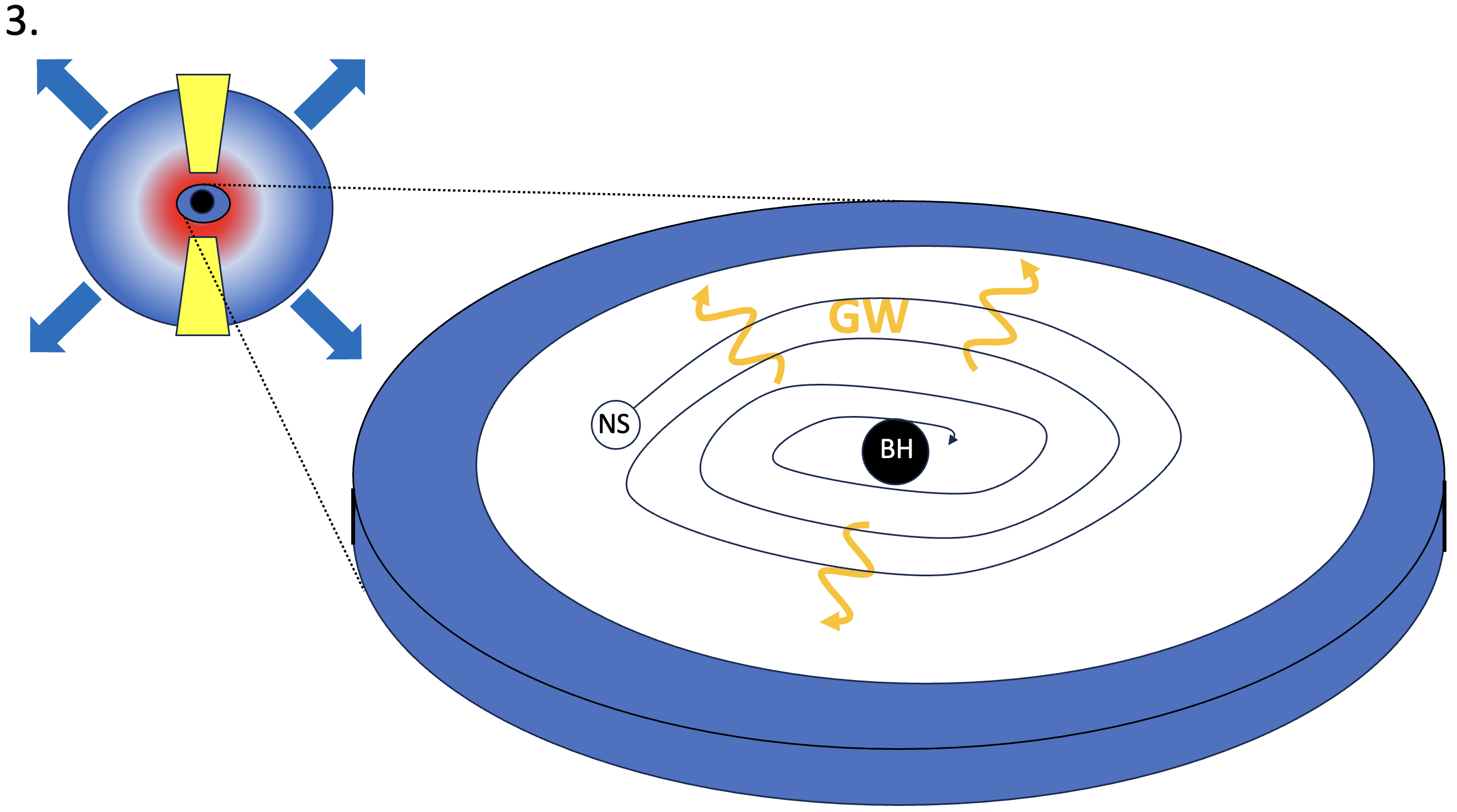}
    \caption{Schematic illustration of the stages of the presented model: (1) core-collapse of a massive rotating star, generating an accretion disk which feeds the central black hole at a high rate $\dot{M} \gtrsim M_{\odot}$ s$^{-1}$. Subsequent fragmentation of the disk at radii $R_{\rm d} \gtrsim 100R_{\rm g}$ where $Q \lesssim Q_0 \sim 1$ leads to the collapse and formation of gravitationally-bound compact objects, potentially including tight binaries containing sub-solar mass neutron stars; (2) binary mergers of the neutron stars formed within the disk lead to LIGO/Virgo-band gravitational wave coalescence events, $r$-process mass ejection, and the creation of energetic compact merger remnants, potentially including accreting black holes and millisecond magnetars (the ordinarily discussed central engines of short GRBs). Along with the disk winds and relativistic jet associated with accretion onto the central black hole, these processes feed energy and $r$-process elements into the unbound ejecta shell from the exploding star; (3) the remaining compact remnant(s), potentially following a chain of hierarchical merging, inspiral into the central black hole, producing a final LIGO/Virgo-band merger.}
    \label{fig:schematic}
\end{figure}

In light of the renewed interest in sub-solar mass compact objects, we revisit the conditions for gravitational instability-induced fragmention and NS formation in collapsar disks. We further consider the possibility that multiple such ssNS, formed, e.g., through the fission of self-gravitating clumps, could merge with one another prior to their coalescence with the central black hole. We highlight several predictions of this channel testable for future gravitational wave events detected by LIGO/Virgo or its successor observatories. The envisioned scenario is summarized in Figure \ref{fig:schematic}, the details of which will be fleshed out as we go along.

\section{Neutron Star Mergers in Collapsar Disks}
\subsection{Gravitationally-Unstable Collapsar Disks}

We consider the core-collapse of a massive $\gtrsim 20M_{\odot}$ star, which fails to explode promptly and results in the formation a black hole of mass $M_{\bullet} \sim 3-30M_{\odot}$. If the star is rotating sufficiently rapidly at the time of collapse, then the in-falling outer layers of its core (i.e., those with the largest angular momentum) will not directly fall into the black hole, but will instead first land in a centrifugally-supported torus, over timescales of seconds to minutes set by the free-fall time of the star. This torus subsequently accretes towards the black hole at a very high rate $\dot{M} \gtrsim 0.1-1 M_{\odot}$ s$^{-1}$, potentially powering a relativistic bipolar jet which breaks out of the star and generates GRB emission (e.g., \citealt{Woosley93,Gottlieb+23}).  

At large radii from the black hole $r \gtrsim 100 R_{\rm g}$, where $R_{\rm g} \equiv GM_{\bullet}/c^{2}$, the collapsar disk can become susceptible to instabilities arising due to self-gravity (e.g., \citealt{Chen&Beloborodov07}). In particular, the wavelength of the fastest growing unstable mode can fit inside the disk for values of the \citet{Toomre64} parameter
\be
Q \equiv \frac{c_{\rm s}\kappa}{\pi G \Sigma} \le Q_0,
\label{eq:Q}
\ee
where $Q_0 \approx 1-2$ (e.g., \citealt{Lodato&Rice04}).  Here, $\Sigma$ and $\kappa \simeq \Omega \approx (GM_{\bullet}/r^{3})^{1/2}$ are the local surface density and epicyclic/orbital frequency of the disk; $c_{\rm s}$ and $h \simeq c_{\rm s}/\Omega \ll r$ are the midplane sound speed and vertical scale-height. Condition \eqref{eq:Q} can be written as a lower limit on the local disk mass:
\be
M_{\rm d} \equiv 2\pi r^{2}\Sigma > \frac{2}{Q_0}\frac{h}{r}M_{\bullet}.
\label{eq:Mdcrit}
\ee
Vertical hydrostatic equilibrium of the disk requires:
\be
h \simeq \frac{2P}{\Sigma \Omega^{2}},
\label{eq:HE}
\ee
where $P(\rho,T,Y_{e}) = \rho c_{\rm s}^{2}$ is the midplane pressure, which depends on the temperature $T$, density $\rho \simeq \Sigma/2h$ and electron fraction $Y_{e}$. The pressure includes contributions from photons, ions, and electrons/positrons; the latter are typically mildly relativistic and mildly degenerate for the conditions of interest.

The midplane pressure, and hence the disk aspect ratio $h/r$, is determined by the balance between heating and cooling. The primary source of heating, at least prior to the formation of any embedded compact objects, is dissipation associated with shear viscosity. This occurs at a specific rate:
\be
\dot{q}_{+} \approx \nu \left(\frac{d\Omega}{d{\rm ln}r}\right)^{2} \approx \frac{9}{4}\alpha r^{2}\Omega^{3}\left(\frac{h}{r}\right)^{2},
\label{eq:qdotplus}
\ee 
where we employ the \citet{Shakura&Sunyaev73} parameterization $\nu = \alpha c_{\rm s}h \approx \alpha r^{2}\Omega(h/r)^{2}$ with $\alpha < 1$ a dimensionless parameter.

While the inner regions of collapsar disks can be opaque to thermal neutrino emission, the disk is transparent to neutrinos at the larger radii where gravitational instabilities operate. The disk is also generally sufficiently hot ($kT \gtrsim 1$ MeV) that heavy elements from the infalling star (e.g., $^{4}$He) at least partially dissociate into free neutrons and protons. The dominant neutrino cooling mechanism under these conditions is the capture of electrons or positrons onto the free nucleons, 
\be
e^{-} + p \rightarrow \nu_{e} + n; \,\,\, e^{+} + n \rightarrow \bar{\nu}_{e} + p,
\label{eq:URCA}
\ee
which provide a total specific cooling rate:
\be
\dot{q}_{-} = \dot{q}_{e^{-}p} + \dot{q}_{e^{+}n}.
\label{eq:qdotminus}
\ee
Additional cooling can occur from helium dissociation \citep{Piro&Pfahl07}, though we neglect this process here for simplicity and return to its effect later on.

The processes \eqref{eq:URCA} not only cool the disk, but can change the ratio of protons to neutrons, i.e. the electron fraction $Y_e$, from the symmetric initial composition $Y_{e,0} \simeq 0.5$ of the infalling stellar material (e.g., \citealt{Siegel+19}). Again neglecting neutrino absorptions, these weak interactions act to drive $Y_{e}$ to an equilibrium value:
\be
Y_{e,eq} \simeq \frac{\lambda_{e^{+}n}}{\lambda_{e^{-}p}+\lambda_{e^{+}n}},
\label{eq:Yeeq}
\ee 
where $\lambda_{\rm e^{-}p}(T,\rho,Y_{e}) \equiv n_{e^{-}}\langle \sigma_{\rm e^{-}p} v_{\rm e^{-}} \rangle$ and $\lambda_{\rm e^{+}n}(T,\rho,Y_{e}) \equiv n_{e^{+}}\langle \sigma_{\rm e^{+}n} v_{\rm e^{+}} \rangle$ are the rates of electron and positron captures (Eq.~\eqref{eq:URCA}), where $\sigma$ are the relevant cross sections. The inner regions of collapsar disks are sufficiently dense that electrons are degenerate; this suppresses positron formation despite the high temperatures $kT \gtrsim 2m_e c^{2}$, such that $\lambda_{e^{+}n} \lesssim \lambda_{e^{-}p}$ and hence $Y_{e,eq} < 0.5$ (e.g., \citealt{Beloborodov03,Metzger+08b}). Such ``neutronization'' of the inflowing gas may have sufficient time to occur provided the electron capture time $t_{\rm n} \equiv \lambda_{e^{-}p}^{-1}$ is shorter than the local accretion time $t_{\rm acc} = M_{\rm d}/\dot{M}$, where $\dot{M} \simeq 3\pi \nu \Sigma$ is the accretion rate.

\subsection{Conditions for Disk Fragmentation and ssNS Formation}

\begin{figure}
    \includegraphics[width=1.0\textwidth]{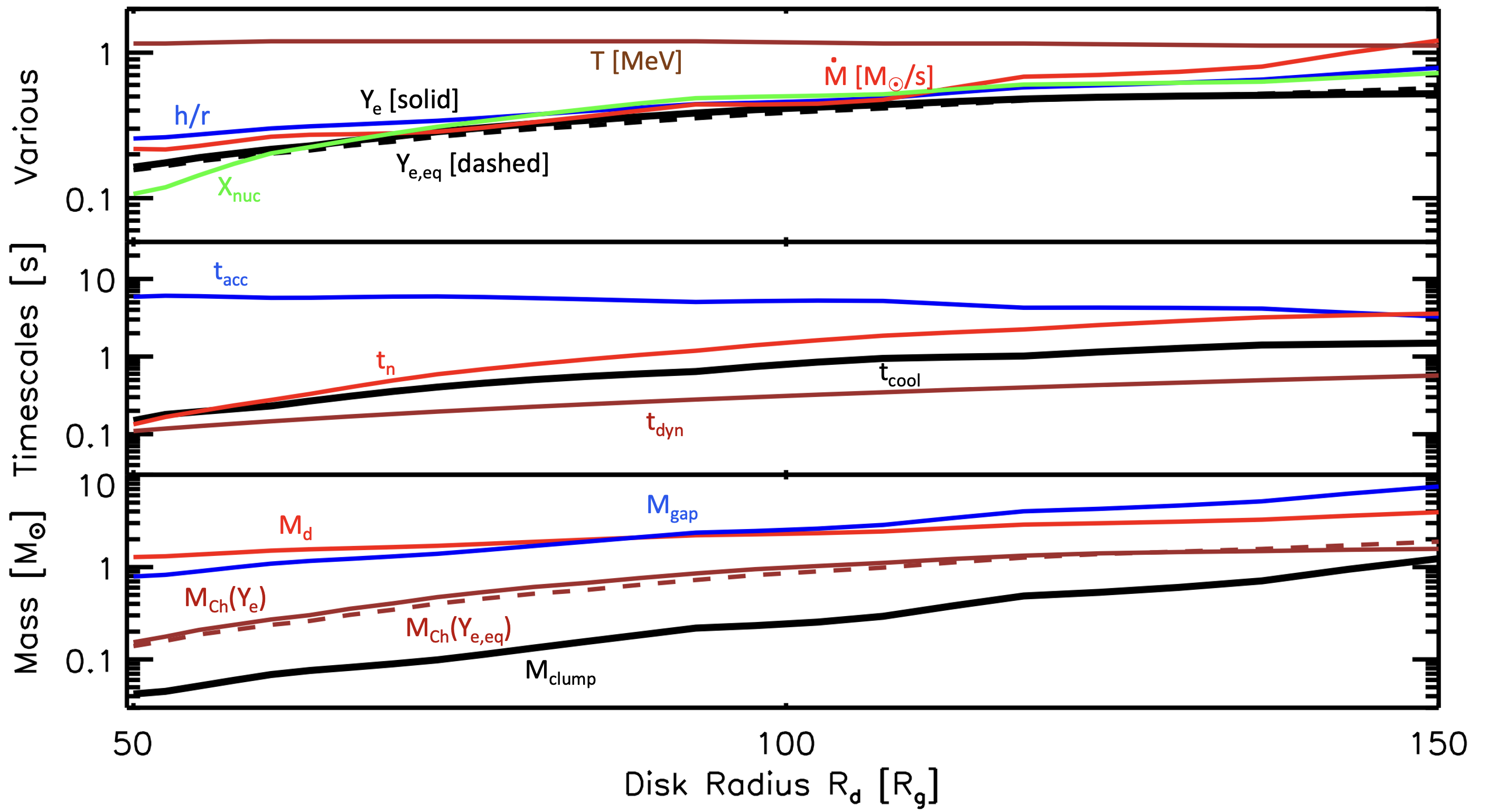}
    \caption{Properties of marginally Toomre-stable ($Q = 1.5 \approx Q_{0}$) collapsar disks as a function of the radius of the disk annulus $r = R_{\rm d}$ for an assumed central black hole mass $M_{\bullet} = 10M_{\odot}$ and viscosity $\alpha = 0.03$. 
    {\it Top Panel:} Estimated electron fraction $Y_e$ after one accretion time (Eq.~\eqref{eq:Ye}; black solid) and equilibrium electron fraction $Y_{e,eq}$ (Eq.~\eqref{eq:Yeeq}; black dashed); disk aspect ratio $h/r$ (blue); midplane temperature $T$ (brown); free nucleon mass-fraction $X_{\rm nuc}$ (green); accretion rate $\dot{M}$ (red); {\it Middle Panel:} neutrino cooling time $t_{\rm cool}$ (black); dynamical/orbital time $t_{\rm dyn} = 2\pi/\Omega$ (brown); neutronization timescale $t_{\rm n} = \lambda_{e^{-}p}^{-1}$ (red); accretion time $t_{\rm acc}$ (blue); {\it Bottom Panel:} local disk mass $M_{\rm d} = 2\pi R_{\rm d}^{2}\Sigma$ (red); characteristic mass of the gravitationally-unstable region $M_{\rm clump} = \pi h^{2}\Sigma$ (black); Chandrasekhar mass $M_{\rm Ch}$, shown separately for the estimated disk electron fraction $Y_e$ (solid brown) and for the equilibrium electron fraction $Y_{e,eq}$ (dashed brown); gap opening mass $M_{\rm gap}$ (blue), above which a bound clump is sufficiently massive to open a gap within the disk.
    }
    \label{fig:diskprofile}
\end{figure}

Figure \ref{fig:diskprofile} shows several of the above-described physical quantities for marginally Toomre-unstable ($Q \approx Q_0 = 1.5$) accretion onto a black hole of mass $M_{\bullet} = 10M_{\odot}$ for an effective viscosity $\alpha = 0.03$ typical of the those provided by the magneto-rotational instability \citep{Davis+10} or the maximum stress provided by gravito-turbulence in gravitationally-unstable disks (\citealt{Gammie01,Rice+05}). For different assumed radii of the disk annulus $r = R_{\rm d}$, we enforce hydrostatic balance (Eq.~\eqref{eq:HE}) and thermal equilibrium ($\dot{q}^{+} = \dot{q}^{-}$;~Eq.~\eqref{eq:qdotplus},\eqref{eq:qdotminus}), using standard expressions for the equation of state, pair-capture rate and associated neutrino cooling rates (e.g., \citealt{Beloborodov03}) for an ideal gas of photons, free nucleons, alpha particles, electrons, and positrons allowing for arbitrary degrees of degeneracy and relativism. 

The disk structure depends on the electron fraction, which in turn depends on the efficacy of weak interactions. Following the discussion at the end of the previous section, we approximate the disk electron fraction as:
\be
Y_e = Y_{e,0}\left(1-e^{-t_{\rm n}/t_{\rm acc}}\right) + Y_{e,eq}e^{-t_{\rm n}/t_{\rm acc}},
\label{eq:Ye}
\ee
where $Y_{e,0} = 0.5$ is the initial electron fraction of the disk from the infalling stellar material and $Y_{e,eq}$ (Eq.~\eqref{eq:Yeeq}) is the equilibrium value attained if the pair-capture reactions \eqref{eq:URCA} come into equilibrium ($t_{\rm n} \ll t_{\rm acc})$.

Fig.~\ref{fig:diskprofile} shows that the collapsar disk must be massive $M_{\rm d} \gtrsim M_{\odot}$ with a high accretion rate $\dot{M} \gtrsim 0.5 M_{\odot}$ s$^{-1}$ to become gravitationally-unstable. These properties are consistent with those achieved by the collapse of rapidly-spinning very-massive stars $\gtrsim 30M_{\odot}$ for which the gravitational free-fall time of the helium core is typically tens of seconds (e.g., \citealt{Siegel+19}). For $R_{\rm d} \sim 100 R_{\rm g}$ the disk is moderately geometrically-thin as a result of neutrino cooling ($h/r \sim 0.3$) with a characteristic midplane temperature $kT \approx 1$ MeV (top panel of Fig.~\ref{fig:diskprofile}). The latter varies only weakly with $R_{\rm d}$ because of the sensitive temperature dependence of the neutrino cooling rate ($\dot{q}_{-} \propto T^{6}$, roughly). The composition is partially free nucleons $X_{\rm nuc} \gtrsim 0.3$, with the remaining mass $X_{\rm He} \simeq 1-X_{\rm nuc}$ mostly in alpha particles. 

Pair-captures favor driving the disk to a neutron-rich composition $Y_{e,eq} < 0.5$ for small disk sizes $R_{\rm d} \lesssim 100 R_{\rm g}$, as a result of the higher electron degeneracy that accompanies the greater density of smaller disks. Furthermore, the timescale for electron captures is sufficiently short relative to the accretion timescale ($t_{\rm n} \lesssim t_{\rm acc}$), that the disk matter has time to become appreciably neutron-rich before flowing inwards to the black hole ($Y_{e} \approx Y_{e,eq} < 0.5$). In fact, the timescale for neutronization is only moderately longer than the dynamical or cooling timescales $t_{\rm n} \sim t_{\rm cool}, t_{\rm dyn}$ for small $R_{\rm d}$. As described below, this has important consequences for the minimum masses of any NS formed.

The disk being gravitationally-unstable does not guarantee that the resulting over-densities will fragment into gravitationally-bound bodies (``clumps''), nor that such clumps would continue to contract to become NS.  For NS formation to occur, at least two conditions must be satisfied:
\begin{enumerate}
    \item 
    
    The proto-clumps must cool radiatively faster than they are sheared apart by the mean flow, i.e. \citep{Gammie01}
\be t_{\rm cool}/t_{\rm dyn} \lesssim \mathcal{O}(1), 
\label{eq:gammie}
\ee
must be satisfied, where $t_{\rm dyn} \equiv 2\pi/\Omega$ and $t_{\rm cool} \equiv e_{\rm th}/(\rho\dot{q}^{-})$ are the dynamical/orbital and radiative cooling time, respectively, and $e_{\rm th}$ is the thermal energy density. The precise order-unity threshold on the right side of \eqref{eq:gammie} is uncertain in the present context, as it depends on the equation of state and cooling function of the disk and must be ascertained with hydrodynamical simulations (e.g., \citealt{Rice+05,Chen+23}).

\item 

To form a NS, the mass of the collapsing self-bound clump must exceed the Chandrasekhar mass (Eq.~\eqref{eq:Mch}), 
i.e.
\be
M_{\rm clump} \gtrsim M_{\rm Ch}(Y_{e}).
\label{eq:McgtMch}
\ee
This way, even once thermal pressure is radiated away, electron degeneracy pressure is unable to support the clump, enabling its continued collapse to nuclear densities. 

\end{enumerate}

The first condition \eqref{eq:gammie} is at best marginally satisfied in the initial disk state for our example solution, for which $t_{\rm cool} \sim t_{\rm dyn}$ (middle panel of Fig.~\ref{fig:diskprofile}). However, since the neutrino-cooling rate increases with temperature ($t_{\rm cool} \propto T^{-2}\rho^{-1}$, roughly) as a marginally-bound region of the disk is compressed to higher densities (roughly adiabatically, $\rho \propto T^{1/3}$), then the ratio $t_{\rm cool}/t_{\rm dyn} \propto \rho^{-11/6}$ becomes shorter as the collapse proceeds, where $t_{\rm dyn} \propto \rho^{-1/2}$ is now the free-fall time of the clump. The additional cooling available from further alpha-particle dissociation during the collapse, of up to $ Q_{\alpha}(1-X_{\rm nuc}) \gtrsim $ 3 MeV per nucleon $\gtrsim kT$ (where $Q_{\alpha} \simeq 7$ MeV per nucleon is the total energy release per dissociated alpha particle), strengthens this conclusion.

Furthermore, in cases where the disk grows via infall from a surrounding stellar envelope (as may describe the collapsar case), the fragmentation criterion may be more nuanced \citep{Kratter&Lodato16}. The maximum level of angular momentum transport (i.e., the effective value of ``$\alpha$'') set by gravito-turbulence depends on the disk properties, which in turn depend on the balance between infall from the stellar envelope adding mass to the disk and accretion depleting it. For an isothermal disk, \citet{Kratter+10} show that fragmentation to form a binary or multiple system occurs above a minimum mass-infall rate, which can be expressed as:
\be
\dot{M} > \dot{M}_{\rm K} \equiv K\frac{(c_{\rm s}^{3}/G)^{5/3}}{(M_{\bullet}\Omega)^{2/3}} \approx 65M_{\odot}\,{\rm s^{-1}}\,\left(\frac{K}{90}\right)\left(\frac{kT}{\rm 1\,MeV}\right)^{5/2}\left(\frac{R_{\rm d}}{100R_{\rm g}}\right),
\ee
where in the second line we have approximated the disk's midplane sound-speed $c_{\rm s} \approx (kT/m_p)^{1/2}$ assuming ion pressure dominates. Applying the prefactor $K \approx 90$ found empirically by \citet{Kratter+10}, the required values $\dot{M}_{\rm K} \sim 10-100 M_{\odot}$ s$^{-1}$ are unphysically large, except perhaps for the most massive collapsing stars $\gtrsim 100 M_{\odot}$ \citep{Siegel+21}. On the other hand, collapsar disks are not isothermal and other physical processes like $\alpha$-particle dissociation cooling may come into play, motivating additional simulation work in the present context to arrive at a definitive conclusion regarding fragmentation.

Regarding the second condition \eqref{eq:McgtMch}, the initial mass of the gravitationally-bound clumps may be crudely estimated by the disk's mass enclosed within a region of radius equal to the maximum Toomre unstable wavelength $\sim h$, i.e.~  
\be
M_{\rm clump} \simeq \pi h^{2}\Sigma.
\label{eq:Mclump}
\ee
Our solutions find $M_{\rm clump} \sim 0.03-1 M_{\odot}$ across the disk radii of interest, typically several times smaller than $M_{\rm ch}$. Although \eqref{eq:McgtMch} is not initially satisfied (in this example), it may become so at a later stage. Firstly, the electron fraction (and hence $M_{\rm Ch}$) will continue to decrease from its initial value at the time of fragmentation, as the temperature of a collapsing clump rises, shortening the timescale for electron captures (roughly as $t_{\rm n} \propto T^{-5}$). 

Secondly, any gravitationally-bound clumps can continue to grow over several orbits, e.g. as a result of mergers with other clumps or potentially gas accretion. A bound object is only sufficiently massive to open a gap in the disk, above a critical mass which is frequently approximated as (e.g., \citealt{Crida+06})
\be
M_{\rm gap} \approx \frac{50 \nu M_{\bullet}}{R_{\rm d}^{2}\Omega}.
\label{eq:Mgap}
\ee 
The fact that $M_{\rm gap} \sim 1-10M_{\odot}$ across the range of $R_{\rm d}$ exceeds the clump mass (Fig.~\ref{fig:diskprofile}) makes gap opening unlikely. This implies that gravitationally-bound clumps can in principle grow through mergers with other clumps located at the same annulus in the disk (or, potentially, through gas accretion) to masses $\gtrsim M_{\rm Ch}$ capable of collapse.  Once forming, however, a NS may not continue to grow appreciably due to the powerful feedback that super-critical accretion has on the growth rate of a compact object (e.g., \citealt{Blandford&Begelman99}).

In summary, while there are many uncertainties, we conclude that collapsar disks are plausibly capable of (a) becoming gravitationally-unstable, (b) fragmenting into bound objects as a result of neutrino-cooling and alpha-dissociation, and (c) producing fragments which either initially, or through subsequent accretion and/or contraction/neutronization, exceed the Chandrasekhar mass $M_{\rm Ch} \propto Y_{e}^{2}$ and thus can collapse to NSs, (d) the latter potentially with sub-solar masses due to the neutron-rich composition of the electron-degenerate disk and fragments, $Y_{e} \lesssim 0.5$. These conditions for creating ssNS are generally satisfied for massive disks $\gg M_{\odot}$ formed with large sizes $R_{\rm d} \sim 100 R_{\rm g}$ (see also \citealt{Piro&Pfahl07}). 

\subsection{Binary NS Formation and Hierarchical Mergers Thereafter}
\label{sec:after}

Once the collapse of a bound clump with mass $\gtrsim M_{\rm Ch}$ is underway, its subsequent evolution will qualitatively resemble the final stages of the core-collapse of a massive star. The collapse process will unfold on the free-fall time of the clump, which equals the disk's dynamical timescale $t_{\rm dyn} \lesssim $ seconds (Fig.~\ref{fig:diskprofile}). The final phase as the proto-NS undergoes Kelvin-Helmholtz contraction to a radius $R_{\rm ns} \lesssim 20$ km takes place on a similar timescale of a few seconds \citep{Burrows&Lattimer86}. 

However, one potentially significant barrier to collapse is the large specific angular momentum of the clump $j_{\rm clump} \sim h (\delta v) \approx h^{2}\Omega$ (where $\delta v \sim h \Omega$ is the velocity shear across the clump), which generally exceeds the maximum rotational angular momentum of a NS, $j_{\rm max} \lesssim (R_{\rm ns} GM_{\rm ns})^{1/2}$ by a factor of several. One way that this barrier could be overcome if a collapsing fragment fissions into one or more sub-bodies, placing its excess angular momentum into orbital motion (see \citealt{Colpi&Rasio94} and \citealt{Colpi&Wasserman02} for related early ideas in the context of NS-mergers and core-collapse supernovae, respectively).\footnote{A similar process has been proposed to form Kuiper Belt binaries such as {\it Ultima Thule} within the Sun's protoplanetary disk (e.g., \citealt{Nesvorny+10}).} Fluid simulations by \citet{Alexander+08} are broadly supportive of this possibility, though additional studies are needed. Equating $j_{\rm clump}$ to the orbital angular momentum $j_{\rm orb} = (2GM_{\rm ns}a_{\rm bin})^{1/2}$ of a binary of two equal mass NS, gives an estimate of the binary separation:
\be
a_{\rm bin} = \frac{R_{\rm d}}{2}\frac{M_{\bullet}}{M_{\rm ns}}\left(\frac{h}{r}\right)^{4} \approx 120\,{\rm km}\,\left(\frac{R_{\rm d}}{100R_{\rm g}}\right)\left(\frac{M_{\rm ns}}{0.5M_{\odot}}\right)^{-1}\left(\frac{M_{\bullet}}{10M_{\odot}}\right)^{2}\left(\frac{M_{\rm d}}{0.3M_{\bullet}}\right)^{4},
\ee
where we have used Eq.~\eqref{eq:Mdcrit} for $Q = Q_0 = 2$. We note that $a_{\rm bin} < r_{\rm H}$ for fiducial parameters, i.e. the putative NS binary fits within its Hill sphere.

Neglecting any additional sources of binary tightening (e.g., due to torques from the surrounding gas disk; \citealt{Stone+17}), the newly-formed NS binary will merge through gravitational-wave emission on a timescale:
\be
\tau_{\rm NS-NS} \simeq \frac{5}{512}\frac{c^{5}a_{\rm bin}^{4}}{G^{3}M_{\rm ns}^{3}} \simeq 17\,{\rm s}\left(\frac{a_{\rm bin}}{120\,{\rm km}}\right)^{4}\left(\frac{M_{\rm ns}}{0.5M_{\odot}}\right)^{-3}.
\ee
Aside from their potential to involve sub-solar mass bodies, this gravitational wave signal may be unique in other ways from most other NS mergers in nature (e.g., those formed through binary star evolution). Insofar that $\tau_{\rm NS-NS}$ is longer than the orbital time of the binary around the central black hole $t_{\rm dyn} \sim 0.1-1$ s (Fig.~\ref{fig:diskprofile}), the early stages of the chirp as viewed by an external observer would be subject to Doppler modulation, at a frequency $f_{\rm orb} \sim 1/t_{\rm dyn} \sim 1-10$ Hz. This effect on the gravitational waveform may be detectable with LIGO/Virgo in some events (e.g., \citealt{Meiron+17}).  We might also expect the frequency of the orbital modulation of the GW-signal to be anti-correlated with the mass of the NS binary, since $M_{\rm clump}$ and $M_{\rm Ch}$ typically increase with $R_{\rm d}$. We also note that if $\tau_{\rm NS-NS}$ is sufficiently short, then the merging proto-neutron stars may still be inflated from their formation process, with radii larger than in their asymptotic cold state; this could improve the prospects for detecting tidal effects in the gravitational waveform (e.g., \citealt{Bandopadhyay+23}).

The end product of the merger of a NS is a compact remnant, either a black hole or NS, depending primarily on the total mass of the binary (e.g., \citealt{Margalit&Metzger19}). Although this merger product will subsequently release enormous energy, mostly in the form of roughly axisymmetric electromagnetic outflows (Sec.~\ref{sec:EM}), neither these$-$nor the recoil kick received from the gravitational waves$-$are likely to be sufficient to unbind the merger product from the potential well of the central black hole (the escape speed at $\lesssim 200 R_{\rm g}$ is $\gtrsim 0.1 c$). This implies that if multiple fragments form and collapse to NS(s) at separate locations in the disk, hierarchical mergers between these objects can take place, similar to those envisioned to occur between binary black holes in dense stellar environments (e.g., \citealt{Gerosa&Berti17}). On the other hand, strong gravitational interactions between compact objects that don't lead to mergers could result in solitary ssNS receiving large kicks that unbind them from the disk. Such a body would only likely be detectable (or, at least, be identifiable as a low-mass NS) if it were to somehow end up in a binary system. Such a match is unlikely to take place in the field, but could in principle take place for core-collapse events in dense stellar environments such as galactic nuclei (e.g., \citealt{Jermyn+22}).

Regardless of the prolificity of collapsar disks, the final products of any such merger chain must eventually merge with the central black hole \citep{Piro&Pfahl07}. If such high mass-ratio inspiral(s), which begin from an initial separation $a_{\rm bin} \sim R_{\rm d} \sim 100 R_{\rm g}$ corresponding to the fragmentation radius, are also driven exclusively by gravitational-waves, then the delay of the merger time with respect to the associated earlier in-disk mergers is given by:
\be
\tau_{\rm BH-NS} \simeq \frac{5}{256}\frac{c^{5}R_{\rm d}^{4}}{G^{3}M_{\bullet}^{2}(2M_{\rm ns})} \approx 10^{3}\,{\rm s}\,\left(\frac{R_{\rm d}}{100R_{\rm g}}\right)^{4}\left(\frac{M_{\rm ns}}{0.5M_{\odot}}\right)^{-1}\left(\frac{M_{\bullet}}{10M_{\odot}}\right)^{2},
\ee 
i.e. minutes to hours, depending sensitively on the details of the system. Type II migration in the gaseous disk may lead to faster migration, causing observable deviation from the vacuum gravitational-wave signal \citep{Piro&Pfahl07}. Aside from the compact merger signals described above, regions of the disk which do not cool efficiently to fragment into bound remnants, can also generate a quasi-periodic gravitational wave signal as the result of spiral density waves (e.g., \citealt{Siegel+21}) or trapped Rossby waves \citep{Gottlieb+24}.

\subsection{Electromagnetic Counterparts: Kilonovae-Embedded Collapsars}
\label{sec:EM}

The above-described processes take place over a window of at most days, embedded within the environment of an exploding star. Even if the initial stellar core-collapse failed to produce a successful explosion, outflows from the black hole accretion disk and relativistic jet are sufficient to power the ejection of several solar masses of material at high velocities $v \gtrsim 0.1 c$ (e.g., \citealt{MacFadyen&Woosley99}), explaining the observed coincidence of energetic (``broad-lined'') supernovae in association with most long GRBs (e.g., \citealt{Woosley&Bloom06}). The discovery of a long GRB or supernova following the gravitational wave trigger from a compact object merger would thus be a smoking gun prediction of the scenario.

The presence of NS merger(s) within the collapsar disk would have several implications for the observable appearance of the associated supernovae. Neutron-rich material released during the merger process create heavy $r$-process elements (e.g., \citealt{Freiburghaus+99}), most of which is ejected from the binary with sufficiently high velocities $\sim 0.1-0.3 c$ to become unbound from the collapsar disk. Radioactive decay within these ejecta will likely not power the cleanly observable kilonova signal seen from normal binary neutron star mergers (\citealt{Metzger+10}) because of the dense, opaque surroundings of the exploding star. However, the presence of heavy $r$-process elements in the collapsar ejecta can still impact the light curves and spectra of the supernova as a result of the high opacities of lanthanide/actinide elements relative to ordinary supernova ejecta (e.g., \citealt{Siegel+19,Barnes&Metzger22,Patel+24}).

Insofar that the merger of two ssNS will create a final remnant of mass $\lesssim 2M_{\odot}$ less than the Tolman-Oppenheimer-Volkoff limit, the merger will form a stable NS rather than a black hole. Such a NS remnant is necessarily very rapidly spinning (e.g., \citealt{Radice+18}) and likely strongly magnetized, i.e., a ``millisecond magnetar'' (e.g., \citealt{Metzger+08a,Combi&Siegel23,Kiuchi24}). After forming, the magnetar will undergo rapid magnetic dipole braking, releasing a large fraction of its $\sim 10^{52}-10^{53}$ ergs of rotational energy into the environment in the form of a magnetized wind over a timescale of minutes to hours (e.g., \citealt{Bucciantini+12,Metzger&Piro14}). Though expanding at relativistic speeds, the magnetar wind will likely be trapped within, and hence share its energy with, the surrounding expanding supernova ejecta on large scales. This could substantially boost the energetics of the merger-embedded supernovae, even compared to the ``hypernovae'' which accompany ordinary (i.e., non-merger hosting) collapsars.

\section{Summary}

The standard of evidence for treating the future detection of sub-solar mass compact objects as evidence for new physics, such as the existence of primordial black holes, must be very high.  On the other hand, the number of plausible astrophysical channels for creating (much less merging) sub-solar mass compact objects is very limited. Motivated by this tension, we have outlined an admittedly speculative scenario for forming and merging ssNS in a single environment: the gaseous accretion disks created by the collapse of massive rotating stars.  The described scenario supports and expands on earlier work by \citet{Piro&Pfahl07}.  

Although our estimates paint a plausible story,
a number of uncertainties remain, particularly with regards to: (a) whether the stripped progenitor stars of collapsars can possess sufficient angular momentum to create massive $\gtrsim M_{\odot}$ disks at large radii $\gtrsim 100 R_{\rm g}$ around the central black hole; (b) whether the criterion for forming gravitationally-bound objects is in fact satisfied by a combination of neutrino and alpha particle dissociation cooling in a full multi-dimensional turbulent disk environment; (c) the resulting mass spectrum of the bound clumps, and whether clump-fissioning or gas-aided capture leads to binary NS formation; (d) the evolution of the disk electron fraction due to pair captures prior and during gravitational collapse, and how this impacts the masses of the NSs that form; (e) feedback effects on the disk mass and energy budget from accretion onto the collapsed remnants. Some of these issues are directly amenable to numerical simulations and will be pursued in future work.

Given the uncertainties, we have emphasized a number of testable predictions of the proposed scenario:
\begin{itemize} 
\item Doppler modulation of the NS-merger gravitational waveform due to the binary's orbital motion around the central black hole. Tidal effects may also be stronger than than predicted by the cold equation of state, due to the inflated radii of newly-formed NS (the cold radii of ssNS and associated tidal deformability being already much larger than ordinary NS; e.g., \citealt{Bandopadhyay+23}).
\item The potential for multiple hierarchical mergers over a short window of minutes to hours. At least one final coalescence event, likely coincident within hours to days, from the merger product and the central black hole is particularly challenging to avoid \citep{Piro&Pfahl07}. Insofar that multiple unrelated mergers within such a short time-frame from the same region of the sky and luminosity distance are likely to be rare, this prediction seems eminently testable with LIGO/Virgo or future more sensitive gravitational wave observatories, particularly those which will provide adequate sky localization.
\item A series of bright electromagnetic counterparts in the form of a gamma-ray burst jet and its multi-wavelength afterglow fed by accretion onto the black hole, followed over weeks to months by a supernova from the disk wind-aided explosion. The explosion may be boosted in its kinetic energy and enriched in $r$-process elements from the embedded NS merger(s), the latter of which can be tested by late-time observations which probe the inner layers of the supernova ejecta (e.g., \citealt{Rastinejad+23,Anand+24}).
\end{itemize}

\begin{acknowledgments}
We thank the reviewer for insightful comments which improved the manuscript. LH thanks Toni Riotto for helpful conversations.  BDM thanks Morgan May for helpful comments on the manuscript. BDM was supported in part by the National Science Foundation (grant No. AST-2009255) and by the NASA Fermi Guest Investigator Program (grant No.~80NSSC22K1574). LH acknowledges support by the DOE DE-SC011941 and a Simons Fellowship in Theoretical Physics. The Flatiron Institute is supported by the Simons Foundation. This research was supported in part by grant no. NSF PHY-2309135 to the Kavli Institute for Theoretical Physics (KITP).
\end{acknowledgments}

\bibliographystyle{aasjournal}

\begin{thebibliography}{}
\expandafter\ifx\csname natexlab\endcsname\relax\def\natexlab#1{#1}\fi
\providecommand{\url}[1]{\href{#1}{#1}}
\providecommand{\dodoi}[1]{doi:~\href{http://doi.org/#1}{\nolinkurl{#1}}}
\providecommand{\doeprint}[1]{\href{http://ascl.net/#1}{\nolinkurl{http://ascl.net/#1}}}
\providecommand{\doarXiv}[1]{\href{https://arxiv.org/abs/#1}{\nolinkurl{https://arxiv.org/abs/#1}}}

\bibitem[{{Abbott} {et~al.}(2018){Abbott}, {Abbott}, {Abbott},
  {et~al.}}]{Abbott_subsolar+18}
{Abbott}, B.~P., {Abbott}, R., {Abbott}, T.~D., {et~al.} 2018, \prl, 121,
  231103, \dodoi{10.1103/PhysRevLett.121.231103}

\bibitem[{{Abbott} {et~al.}(2022)}]{Abbott_subsolar_22}
{Abbott}, R., {et~al.} 2022, \prl, 129, 061104,
  \dodoi{10.1103/PhysRevLett.129.061104}

\bibitem[{{Alexander} {et~al.}(2008){Alexander}, {Armitage}, \&
  {Cuadra}}]{Alexander+08}
{Alexander}, R.~D., {Armitage}, P.~J., \& {Cuadra}, J. 2008, \mnras, 389, 1655,
  \dodoi{10.1111/j.1365-2966.2008.13706.x}

\bibitem[{{Anand} {et~al.}(2024){Anand}, {Barnes}, {Yang}, {Kasliwal},
  {Coughlin}, {et~al.}}]{Anand+24}
{Anand}, S., {Barnes}, J., {Yang}, S., {et~al.} 2024, \apj, 962, 68,
  \dodoi{10.3847/1538-4357/ad11df}

\bibitem[{{Antoni} \& {Quataert}(2023)}]{Antoni&Quataert23}
{Antoni}, A., \& {Quataert}, E. 2023, \mnras, 525, 1229,
  \dodoi{10.1093/mnras/stad2328}

\bibitem[{{Bandopadhyay} {et~al.}(2023){Bandopadhyay}, {Reed}, {Padamata},
  {Leon}, {Horowitz}, {Brown}, {Radice}, {Fattoyev}, \&
  {Piekarewicz}}]{Bandopadhyay+23}
{Bandopadhyay}, A., {Reed}, B., {Padamata}, S., {et~al.} 2023, \prd, 107,
  103012, \dodoi{10.1103/PhysRevD.107.103012}

\bibitem[{{Barnes} \& {Metzger}(2022)}]{Barnes&Metzger22}
{Barnes}, J., \& {Metzger}, B.~D. 2022, \apjl, 939, L29,
  \dodoi{10.3847/2041-8213/ac9b41}

\bibitem[{{Beloborodov}(2003)}]{Beloborodov03}
{Beloborodov}, A.~M. 2003, \apj, 588, 931, \dodoi{10.1086/374217}

\bibitem[{{Blandford} \& {Begelman}(1999)}]{Blandford&Begelman99}
{Blandford}, R.~D., \& {Begelman}, M.~C. 1999, \mnras, 303, L1,
  \dodoi{10.1046/j.1365-8711.1999.02358.x}

\bibitem[{{Boss}(1997)}]{Boss97}
{Boss}, A.~P. 1997, Science, 276, 1836, \dodoi{10.1126/science.276.5320.1836}

\bibitem[{{Bucciantini} {et~al.}(2012){Bucciantini}, {Metzger}, {Thompson}, \&
  {Quataert}}]{Bucciantini+12}
{Bucciantini}, N., {Metzger}, B.~D., {Thompson}, T.~A., \& {Quataert}, E. 2012,
  \mnras, 419, 1537, \dodoi{10.1111/j.1365-2966.2011.19810.x}

\bibitem[{{Burrows} \& {Lattimer}(1986)}]{Burrows&Lattimer86}
{Burrows}, A., \& {Lattimer}, J.~M. 1986, \apj, 307, 178,
  \dodoi{10.1086/164405}

\bibitem[{{Burrows} {et~al.}(2019){Burrows}, {Radice}, \&
  {Vartanyan}}]{Burrows+19}
{Burrows}, A., {Radice}, D., \& {Vartanyan}, D. 2019, \mnras, 485, 3153,
  \dodoi{10.1093/mnras/stz543}

\bibitem[{{Burrows} \& {Vartanyan}(2021)}]{Burrows&Vartanyan21}
{Burrows}, A., \& {Vartanyan}, D. 2021, \nat, 589, 29,
  \dodoi{10.1038/s41586-020-03059-w}

\bibitem[{{Cantiello} {et~al.}(2007){Cantiello}, {Yoon}, {Langer}, \&
  {Livio}}]{Cantiello+07}
{Cantiello}, M., {Yoon}, S.~C., {Langer}, N., \& {Livio}, M. 2007, \aap, 465,
  L29, \dodoi{10.1051/0004-6361:20077115}

\bibitem[{{Carr} {et~al.}(2021){Carr}, {Kohri}, {Sendouda}, \&
  {Yokoyama}}]{Carr+21}
{Carr}, B., {Kohri}, K., {Sendouda}, Y., \& {Yokoyama}, J. 2021, Reports on
  Progress in Physics, 84, 116902, \dodoi{10.1088/1361-6633/ac1e31}

\bibitem[{{Chandrasekhar}(1931)}]{Chandrasekhar31}
{Chandrasekhar}, S. 1931, \apj, 74, 81, \dodoi{10.1086/143324}

\bibitem[{{Chen} \& {Beloborodov}(2007)}]{Chen&Beloborodov07}
{Chen}, W.-X., \& {Beloborodov}, A.~M. 2007, \apj, 657, 383,
  \dodoi{10.1086/508923}

\bibitem[{{Chen} {et~al.}(2023){Chen}, {Jiang}, {Goodman}, \&
  {Ostriker}}]{Chen+23}
{Chen}, Y.-X., {Jiang}, Y.-F., {Goodman}, J., \& {Ostriker}, E.~C. 2023, \apj,
  948, 120, \dodoi{10.3847/1538-4357/acc023}

\bibitem[{{Colpi} \& {Rasio}(1994)}]{Colpi&Rasio94}
{Colpi}, M., \& {Rasio}, F.~A. 1994, \memsai, 65, 379

\bibitem[{{Colpi} \& {Wasserman}(2002)}]{Colpi&Wasserman02}
{Colpi}, M., \& {Wasserman}, I. 2002, \apj, 581, 1271, \dodoi{10.1086/344405}

\bibitem[{{Combi} \& {Siegel}(2023)}]{Combi&Siegel23}
{Combi}, L., \& {Siegel}, D.~M. 2023, \prl, 131, 231402,
  \dodoi{10.1103/PhysRevLett.131.231402}

\bibitem[{Crescimbeni {et~al.}(2024)Crescimbeni, Franciolini, Pani, \&
  Riotto}]{Crescimbeni+24}
Crescimbeni, F., Franciolini, G., Pani, P., \& Riotto, A. 2024, Phys. Rev. D,
  109, 124063, \dodoi{10.1103/PhysRevD.109.124063}

\bibitem[{{Crida} {et~al.}(2006){Crida}, {Morbidelli}, \& {Masset}}]{Crida+06}
{Crida}, A., {Morbidelli}, A., \& {Masset}, F. 2006, \icarus, 181, 587,
  \dodoi{10.1016/j.icarus.2005.10.007}

\bibitem[{{Davis} {et~al.}(2010){Davis}, {Stone}, \& {Pessah}}]{Davis+10}
{Davis}, S.~W., {Stone}, J.~M., \& {Pessah}, M.~E. 2010, \apj, 713, 52,
  \dodoi{10.1088/0004-637X/713/1/52}

\bibitem[{{Doroshenko} {et~al.}(2022){Doroshenko}, {Suleimanov},
  {P{\"u}hlhofer}, \& {Santangelo}}]{Doroshenko+22}
{Doroshenko}, V., {Suleimanov}, V., {P{\"u}hlhofer}, G., \& {Santangelo}, A.
  2022, Nature Astronomy, 6, 1444, \dodoi{10.1038/s41550-022-01800-1}

\bibitem[{{Ertl} {et~al.}(2020){Ertl}, {Woosley}, {Sukhbold}, \&
  {Janka}}]{Ertl+20}
{Ertl}, T., {Woosley}, S.~E., {Sukhbold}, T., \& {Janka}, H.~T. 2020, \apj,
  890, 51, \dodoi{10.3847/1538-4357/ab6458}

\bibitem[{{Freiburghaus} {et~al.}(1999){Freiburghaus}, {Rosswog}, \&
  {Thielemann}}]{Freiburghaus+99}
{Freiburghaus}, C., {Rosswog}, S., \& {Thielemann}, F. 1999, ApJ, 525, L121,
  \dodoi{10.1086/312343}

\bibitem[{{Fuller} {et~al.}(2019){Fuller}, {Piro}, \& {Jermyn}}]{Fuller+19}
{Fuller}, J., {Piro}, A.~L., \& {Jermyn}, A.~S. 2019, \mnras, 485, 3661,
  \dodoi{10.1093/mnras/stz514}

\bibitem[{{Gammie}(2001)}]{Gammie01}
{Gammie}, C.~F. 2001, \apj, 553, 174, \dodoi{10.1086/320631}

\bibitem[{{Gerosa} \& {Berti}(2017)}]{Gerosa&Berti17}
{Gerosa}, D., \& {Berti}, E. 2017, \prd, 95, 124046,
  \dodoi{10.1103/PhysRevD.95.124046}

\bibitem[{{Goodman} \& {Tan}(2004)}]{Goodman&Tan04}
{Goodman}, J., \& {Tan}, J.~C. 2004, \apj, 608, 108, \dodoi{10.1086/386360}

\bibitem[{{Gottlieb} {et~al.}(2023){Gottlieb}, {Jacquemin-Ide}, {Lowell},
  {Tchekhovskoy}, \& {Ramirez-Ruiz}}]{Gottlieb+23}
{Gottlieb}, O., {Jacquemin-Ide}, J., {Lowell}, B., {Tchekhovskoy}, A., \&
  {Ramirez-Ruiz}, E. 2023, \apjl, 952, L32, \dodoi{10.3847/2041-8213/ace779}

\bibitem[{{Gottlieb} {et~al.}(2024){Gottlieb}, {Levinson}, \&
  {Levin}}]{Gottlieb+24}
{Gottlieb}, O., {Levinson}, A., \& {Levin}, Y. 2024, arXiv e-prints,
  arXiv:2406.19452.
\newblock \doarXiv{2406.19452}

\bibitem[{{Jermyn} {et~al.}(2022){Jermyn}, {Dittmann}, {McKernan}, {Ford}, \&
  {Cantiello}}]{Jermyn+22}
{Jermyn}, A.~S., {Dittmann}, A.~J., {McKernan}, B., {Ford}, K.~E.~S., \&
  {Cantiello}, M. 2022, \apj, 929, 133, \dodoi{10.3847/1538-4357/ac5d40}

\bibitem[{{Kiuchi} {et~al.}(2024){Kiuchi}, {Reboul-Salze}, {Shibata}, \&
  {Sekiguchi}}]{Kiuchi24}
{Kiuchi}, K., {Reboul-Salze}, A., {Shibata}, M., \& {Sekiguchi}, Y. 2024,
  Nature Astronomy, 8, 298, \dodoi{10.1038/s41550-024-02194-y}

\bibitem[{{Kratter} \& {Lodato}(2016)}]{Kratter&Lodato16}
{Kratter}, K., \& {Lodato}, G. 2016, \araa, 54, 271,
  \dodoi{10.1146/annurev-astro-081915-023307}

\bibitem[{{Kratter} {et~al.}(2010){Kratter}, {Matzner}, {Krumholz}, \&
  {Klein}}]{Kratter+10}
{Kratter}, K.~M., {Matzner}, C.~D., {Krumholz}, M.~R., \& {Klein}, R.~I. 2010,
  \apj, 708, 1585, \dodoi{10.1088/0004-637X/708/2/1585}

\bibitem[{{Lattimer} \& {Prakash}(2004)}]{Lattimer&Prakash04}
{Lattimer}, J.~M., \& {Prakash}, M. 2004, Science, 304, 536,
  \dodoi{10.1126/science.1090720}

\bibitem[{{Levin}(2003)}]{Levin03}
{Levin}, Y. 2003, arXiv e-prints, astro,
  \dodoi{10.48550/arXiv.astro-ph/0307084}

\bibitem[{{Lodato} \& {Rice}(2004)}]{Lodato&Rice04}
{Lodato}, G., \& {Rice}, W.~K.~M. 2004, \mnras, 351, 630,
  \dodoi{10.1111/j.1365-2966.2004.07811.x}

\bibitem[{{LVK Collaboration}(2023)}]{LVK_23}
{LVK Collaboration}. 2023, \mnras, 526, 6234, \dodoi{10.1093/mnras/stad3120}

\bibitem[{{MacFadyen} \& {Woosley}(1999)}]{MacFadyen&Woosley99}
{MacFadyen}, A.~I., \& {Woosley}, S.~E. 1999, \apj, 524, 262,
  \dodoi{10.1086/307790}

\bibitem[{{Mandel} \& {Broekgaarden}(2022)}]{Mandel&Broekgaarden22}
{Mandel}, I., \& {Broekgaarden}, F.~S. 2022, Living Reviews in Relativity, 25,
  1, \dodoi{10.1007/s41114-021-00034-3}

\bibitem[{{Margalit} \& {Metzger}(2019)}]{Margalit&Metzger19}
{Margalit}, B., \& {Metzger}, B.~D. 2019, \apjl, 880, L15,
  \dodoi{10.3847/2041-8213/ab2ae2}

\bibitem[{{Martinez} {et~al.}(2015){Martinez}, {Stovall}, {Freire}, {Deneva},
  {Jenet}, {McLaughlin}, {Bagchi}, {Bates}, \& {Ridolfi}}]{Martinez+15}
{Martinez}, J.~G., {Stovall}, K., {Freire}, P.~C.~C., {et~al.} 2015, \apj, 812,
  143, \dodoi{10.1088/0004-637X/812/2/143}

\bibitem[{{Meiron} {et~al.}(2017){Meiron}, {Kocsis}, \& {Loeb}}]{Meiron+17}
{Meiron}, Y., {Kocsis}, B., \& {Loeb}, A. 2017, \apj, 834, 200,
  \dodoi{10.3847/1538-4357/834/2/200}

\bibitem[{{Metzger} {et~al.}(2010){Metzger}, {Arcones}, {Quataert}, \&
  {Mart{\'{\i}}nez-Pinedo}}]{Metzger+10}
{Metzger}, B.~D., {Arcones}, A., {Quataert}, E., \& {Mart{\'{\i}}nez-Pinedo},
  G. 2010, \mnras, 402, 2771, \dodoi{10.1111/j.1365-2966.2009.16107.x}

\bibitem[{{Metzger} \& {Piro}(2014)}]{Metzger&Piro14}
{Metzger}, B.~D., \& {Piro}, A.~L. 2014, \mnras, 439, 3916,
  \dodoi{10.1093/mnras/stu247}

\bibitem[{{Metzger} {et~al.}(2008{\natexlab{a}}){Metzger}, {Quataert}, \&
  {Thompson}}]{Metzger+08a}
{Metzger}, B.~D., {Quataert}, E., \& {Thompson}, T.~A. 2008{\natexlab{a}},
  \mnras, 385, 1455, \dodoi{10.1111/j.1365-2966.2008.12923.x}

\bibitem[{{Metzger} {et~al.}(2008{\natexlab{b}}){Metzger}, {Thompson}, \&
  {Quataert}}]{Metzger+08b}
{Metzger}, B.~D., {Thompson}, T.~A., \& {Quataert}, E. 2008{\natexlab{b}},
  \apj, 676, 1130, \dodoi{10.1086/526418}

\bibitem[{{Morr{\'a}s} {et~al.}(2023){Morr{\'a}s}, {Nu{\~n}o Siles},
  {Garc{\'\i}a-Bellido}, {Ruiz Morales}, {Men{\'e}ndez-V{\'a}zquez},
  {Karathanasis}, {Martinovic}, {Phukon}, {Clesse}, {Mart{\'\i}nez}, \&
  {Sakellariadou}}]{Morras+23}
{Morr{\'a}s}, G., {Nu{\~n}o Siles}, J.~F., {Garc{\'\i}a-Bellido}, J., {et~al.}
  2023, Physics of the Dark Universe, 42, 101285,
  \dodoi{10.1016/j.dark.2023.101285}

\bibitem[{{Nesvorn{\'y}} {et~al.}(2010){Nesvorn{\'y}}, {Youdin}, \&
  {Richardson}}]{Nesvorny+10}
{Nesvorn{\'y}}, D., {Youdin}, A.~N., \& {Richardson}, D.~C. 2010, \aj, 140,
  785, \dodoi{10.1088/0004-6256/140/3/785}

\bibitem[{{Nomoto} \& {Kondo}(1991)}]{Nomoto&Kondo91}
{Nomoto}, K., \& {Kondo}, Y. 1991, \apjl, 367, L19, \dodoi{10.1086/185922}

\bibitem[{{Patel} {et~al.}(2024){Patel}, {Goldberg}, {Renzo}, \&
  {Metzger}}]{Patel+24}
{Patel}, A., {Goldberg}, J.~A., {Renzo}, M., \& {Metzger}, B.~D. 2024, arXiv
  e-prints, arXiv:2401.13035, \dodoi{10.48550/arXiv.2401.13035}

\bibitem[{{Piro} \& {Pfahl}(2007)}]{Piro&Pfahl07}
{Piro}, A.~L., \& {Pfahl}, E. 2007, \apj, 658, 1173, \dodoi{10.1086/511672}

\bibitem[{{Popov} {et~al.}(2007){Popov}, {Blaschke}, {Grigorian}, \&
  {Prokhorov}}]{Popov+07}
{Popov}, S., {Blaschke}, D., {Grigorian}, H., \& {Prokhorov}, M. 2007, \apss,
  308, 381, \dodoi{10.1007/s10509-007-9335-9}

\bibitem[{{Radice} {et~al.}(2018){Radice}, {Perego}, {Bernuzzi}, \&
  {Zhang}}]{Radice+18}
{Radice}, D., {Perego}, A., {Bernuzzi}, S., \& {Zhang}, B. 2018, \mnras, 481,
  3670, \dodoi{10.1093/mnras/sty2531}

\bibitem[{{Rastinejad} {et~al.}(2023){Rastinejad}, {Fong}, {Levan}, {Tanvir},
  {Kilpatrick}, {Fruchter}, {Anand}, {Bhirombhakdi}, {Covino}, {Fynbo},
  {Halevi}, {Hartmann}, {Heintz}, {Izzo}, {Jakobsson}, {Lamb}, {Malesani},
  {Melandri}, {Metzger}, {Milvang-Jensen}, {Pian}, {Pugliese}, {Rossi},
  {Siegel}, {Singh}, \& {Stratta}}]{Rastinejad+23}
{Rastinejad}, J.~C., {Fong}, W., {Levan}, A.~J., {et~al.} 2023, arXiv e-prints,
  arXiv:2312.04630, \dodoi{10.48550/arXiv.2312.04630}

\bibitem[{{Rice} {et~al.}(2005){Rice}, {Lodato}, \& {Armitage}}]{Rice+05}
{Rice}, W.~K.~M., {Lodato}, G., \& {Armitage}, P.~J. 2005, \mnras, 364, L56,
  \dodoi{10.1111/j.1745-3933.2005.00105.x}

\bibitem[{{Shakura} \& {Sunyaev}(1973)}]{Shakura&Sunyaev73}
{Shakura}, N.~I., \& {Sunyaev}, R.~A. 1973, \aap, 500, 33

\bibitem[{{Siegel} {et~al.}(2021){Siegel}, {Agarwal}, {Barnes}, {Metzger},
  {Renzo}, \& {Villar}}]{Siegel+21}
{Siegel}, D.~M., {Agarwal}, A., {Barnes}, J., {et~al.} 2021, arXiv e-prints,
  arXiv:2111.03094.
\newblock \doarXiv{2111.03094}

\bibitem[{{Siegel} {et~al.}(2019){Siegel}, {Barnes}, \& {Metzger}}]{Siegel+19}
{Siegel}, D.~M., {Barnes}, J., \& {Metzger}, B.~D. 2019, \nat, 569, 241,
  \dodoi{10.1038/s41586-019-1136-0}

\bibitem[{{Silva} {et~al.}(2016){Silva}, {Sotani}, \& {Berti}}]{Silva+16}
{Silva}, H.~O., {Sotani}, H., \& {Berti}, E. 2016, \mnras, 459, 4378,
  \dodoi{10.1093/mnras/stw969}

\bibitem[{{Stone} {et~al.}(2017){Stone}, {Metzger}, \& {Haiman}}]{Stone+17}
{Stone}, N.~C., {Metzger}, B.~D., \& {Haiman}, Z. 2017, \mnras, 464, 946,
  \dodoi{10.1093/mnras/stw2260}

\bibitem[{{Sukhbold} {et~al.}(2016){Sukhbold}, {Ertl}, {Woosley}, {Brown}, \&
  {Janka}}]{Sukhbold+16}
{Sukhbold}, T., {Ertl}, T., {Woosley}, S.~E., {Brown}, J.~M., \& {Janka}, H.~T.
  2016, \apj, 821, 38, \dodoi{10.3847/0004-637X/821/1/38}

\bibitem[{{Suwa} {et~al.}(2018){Suwa}, {Yoshida}, {Shibata}, {Umeda}, \&
  {Takahashi}}]{Suwa+18}
{Suwa}, Y., {Yoshida}, T., {Shibata}, M., {Umeda}, H., \& {Takahashi}, K. 2018,
  \mnras, 481, 3305, \dodoi{10.1093/mnras/sty2460}

\bibitem[{{Tauris} \& {Janka}(2019)}]{Tauris&Janka19}
{Tauris}, T.~M., \& {Janka}, H.-T. 2019, \apjl, 886, L20,
  \dodoi{10.3847/2041-8213/ab5642}

\bibitem[{{Toomre}(1964)}]{Toomre64}
{Toomre}, A. 1964, \apj, 139, 1217, \dodoi{10.1086/147861}

\bibitem[{{Woosley}(1993)}]{Woosley93}
{Woosley}, S.~E. 1993, \apj, 405, 273, \dodoi{10.1086/172359}

\bibitem[{{Woosley} \& {Bloom}(2006)}]{Woosley&Bloom06}
{Woosley}, S.~E., \& {Bloom}, J.~S. 2006, ARAA, 44, 507,
  \dodoi{10.1146/annurev.astro.43.072103.150558}

\bibitem[{{Woosley} {et~al.}(2020){Woosley}, {Sukhbold}, \&
  {Janka}}]{Woosley+20}
{Woosley}, S.~E., {Sukhbold}, T., \& {Janka}, H.~T. 2020, \apj, 896, 56,
  \dodoi{10.3847/1538-4357/ab8cc1}

\end{thebibliography}

\end{document}